\documentclass[aps,superscriptaddress,nofootinbib,floats]{revtex4}
\usepackage{graphicx}
\usepackage{latexsym}

\begin{document}

\title{MEASUREMENT OF THE CONDUCTANCE OF A HYDROGEN MOLECULE}
\author{R.H.M. Smit}
\affiliation{Kamerlingh Onnes Laboratorium, Universiteit Leiden,
PO Box 9504, NL-2300 RA Leiden, The Netherlands}
\author{Y. Noat}
\altaffiliation[Present address:]{Université Denis Diderot (Paris
7), Groupe de Physique des Solides UMR 75 88, 2 place Jussieu,
75251 Paris Cedex 05, France.}
\affiliation{Kamerlingh Onnes
Laboratorium, Universiteit Leiden, PO Box 9504, NL-2300 RA Leiden,
The Netherlands}
\author{C. Untiedt}
\affiliation{Kamerlingh Onnes Laboratorium, Universiteit Leiden,
PO Box 9504, NL-2300 RA Leiden, The Netherlands}
\author{N. D. Lang}
\affiliation{IBM Research Division, Thomas J. Watson Research
Center, Yorktown Heights, New York 10598, USA}
\author{M. van Hemert}
\affiliation{Gorlaeus Laboratorium, Universiteit Leiden, PO Box
9504, 2300 RA Leiden, The Netherlands}
\author{J.M. van Ruitenbeek}
\email[Correspondence should be addressed to:
]{ruitenbe@phys.leidenuniv.nl} \affiliation{Kamerlingh Onnes
Laboratorium, Universiteit Leiden, PO Box 9504, NL-2300 RA Leiden,
The Netherlands}

\date{\today}

\maketitle

{\bf Recent years have shown steady progress in research towards
molecular electronics \cite{aviram98,langlais99} where molecules
have been investigated as switches \cite{gao00,collier99,reed01},
diodes \cite{metzger98}, and electronic mixers \cite{chen-j99}. In
much of the previous work a Scanning Tunnelling Microscope was
employed to address an individual molecule. As this arrangement
does not provide long-term stability, more recently
metal-molecule-metal links have been made using break junction
devices \cite{reed97,kergueris99,reichert02}. However, it has been
difficult to establish unambiguously that a single molecule forms
the contact \cite{emberly01}.  Here, we show that a single H$_2$
molecule can form a stable bridge between Pt electrodes. In
contrast to results for other organic molecules, the bridge has a
nearly perfect conductance of one quantum unit, carried by a
single channel.  The H$_2$-bridge provides a simple test system
and a fundamental step towards understanding transport properties
of single-molecule devices.}

\begin{figure}[!t]
\begin{center}
\includegraphics[width=180mm]{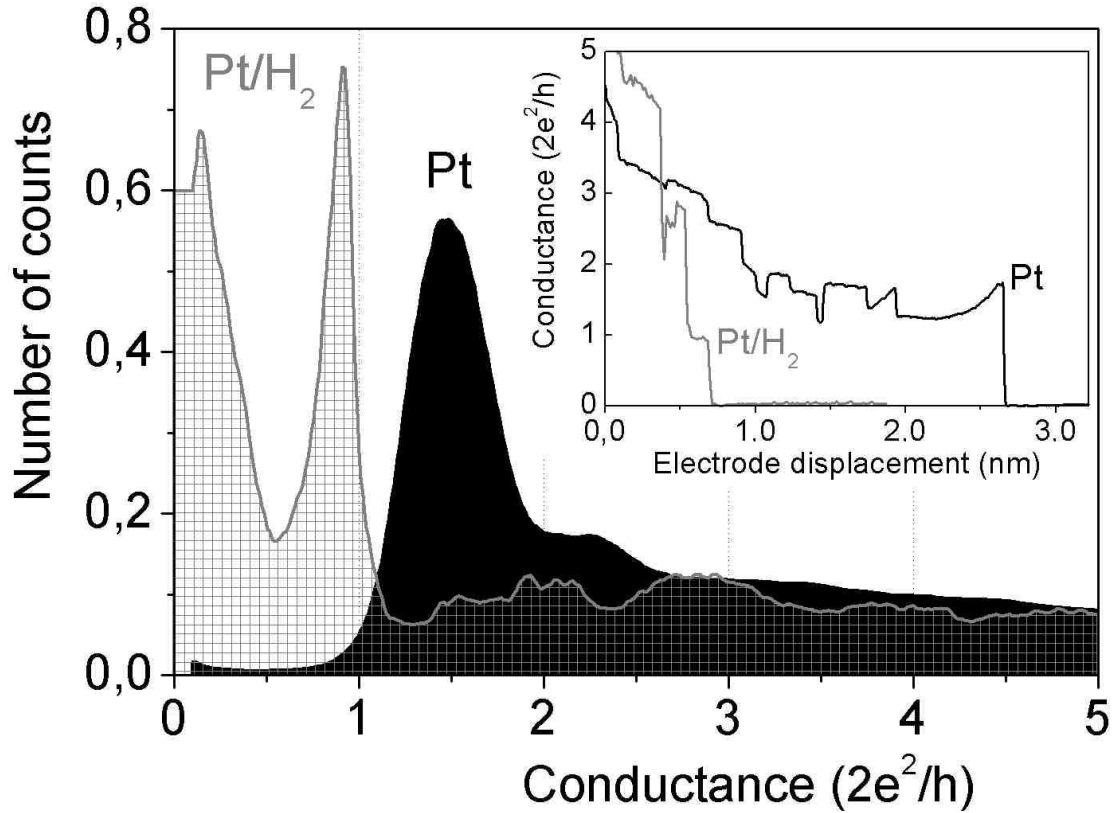}
\end{center}
\caption{Conductance curves and histograms for clean Pt, and Pt in
a H$_2$ atmosphere. The inset shows a conductance curve for clean
Pt (black) at 4.2\,K recorded with a bias voltage of 10\,mV,
before admitting H$_2$ gas into the system. About 10\,000 similar
curves are used to build the conductance histogram shown in the
main panel (black), which has been normalized by the area under
the curve. After introducing hydrogen gas the conductance curves
change qualitatively as illustrated by the grey curve in the
inset, recorded at 100\,mV. This is most clearly brought out by
the conductance histogram (grey; recorded with 140\,mV bias).
Briefly the mechanically controllable break junction technique
works as follows: Starting with a macroscopic metal wire a notch
is formed by incision with a knife. The samples are mounted inside
a vacuum container and pumped to a pressure below $5 \cdot
10^{-7}$\,mbar. Next, the system is cooled to 4.2\,K in order to
attain a cryogenic vacuum. After cooling, the sample wire is
broken at the notch by bending of the substrate onto which it has
been fixed. The clean, freshly exposed fracture surfaces are then
brought back into contact by slightly relaxing the bending. With
the use of a piezoelectric element the displacement of the two
electrodes can be finely adjusted to form a stable contact of
atomic size. A thick copper finger provides thermal contact to the
sample inside the container.} \label{histograms}
\end{figure}

In this work we employ a mechanically controllable break junction
\cite{muller92a,ruitenbeek97a} at low temperatures (4.2\,K) to
produce pure metallic contacts of atomic size. The inset in
Fig.\,\ref{histograms} shows a typical conductance curve for a
clean Pt contact\footnote{We concentrate on results obtained for
Pt wires, but conductance histograms suggest similar behaviour for
Pd.} that is recorded while gradually decreasing the contact size
by ramping up the piezovoltage (black curve). The conductance is
expressed in terms of the quantum unit, $G_0 = 2e^2/h$ , with $e$
the electron charge and $h$ Planck's constant. The jumps in the
conductance are the result of sudden atomic rearrangements in
response to the applied strain \cite{rubio96}. After the
conductance has dropped to a value corresponding to a single atom,
which for Pt is in the range 1.2--2.3\,$G_0$, the contact suddenly
breaks.  In order to extract the common features in these
conductance curves the data of a large series of conductance
curves are collected into a conductance histogram. The main panel
in Fig.\,\ref{histograms} shows such a histogram for Pt contacts
(black). It is dominated by a large peak at 1.4--1.8\,$G_0$, which
represents the range of conductance values for contacts having a
single atom in cross section (for a recent review see
Ref.\,\cite{agrait02b}). The histogram drops sharply to zero for
lower conductance values, as is typically found for Pt contacts in
absence of adsorbates and impurities. Remarkably, the character of
the conductance curves and the shape of the resulting histogram
change dramatically when a small quantity of hydrogen gas is
admitted to the vacuum pot (grey curves). The critical amount is
difficult to establish since most hydrogen is expected to condense
on the walls of the container,\footnote{The equilibrium H$_2$ gas
pressure at a temperature of 4.2\,K is about $10^{-6}$\,mbar.} but
the results are not very sensitive to the precise quantity. The
peak at the position characteristic for Pt disappears and a large
weight is added in the entire range below that value. On top of
this background a distinct peak close to 1\,$G_0$ is found, which
grows for larger currents through the contact, while the
background is suppressed. For still larger bias voltages, above
200\,mV, we recover the histogram for clean Pt. The
low-conductance tail and the peak at 1\,$G_0$ reappear upon
lowering the bias again. The bias dependence of the histograms is
attributed to local heating by the current. For moderate bias the
weakly bound physisorbed H$_2$ is evaporated, and only above
200\,mV the chemically bonded hydrogen molecules are removed from
the Pt surface.

We concentrate now on the remarkable fact that in the presence of
hydrogen there appears to be a frequently occurring stable
configuration that has a conductance of nearly unity, an example
of which is shown in the inset of Fig.\,\ref{histograms}, and
which is responsible for the sharp peak just below 1\,$G_0$ in the
histogram. We can select this configuration by recording
conductance traces of the type shown in the inset, and stop the
motion of the electrodes as soon as we find a plateau at
~1\,$G_0$. In order to investigate the structure of the contact on
this plateau we use point contact spectroscopy (PCS). PCS was
originally developed by Yanson \cite{yanson74,jansen80} for
contacts that are large compared to the atomic scale. One measures
the differential conductance, $G_{\rm d}=dI/dV$ , as a function of
the dc bias voltage. The electrons in the contact are accelerated
to an excess energy $eV$. When this energy reaches that of the
main phonon modes of the metal inelastic scattering results in an
enhanced probability for the electrons to scatter back through the
contact, which is seen as a drop in  $G_{\rm d}$. Recently, this
principle has been applied to single-atom metallic contacts and
chains of atoms \cite{untiedt00,agrait02a}, and a theoretical
description for localised vibration modes in quantum point
contacts has been developed \cite{bonca95,emberly99a}. Vibration
modes for individual molecules have been observed before, using
inelastic electron tunnelling spectroscopy (IETS) in a
low-temperature STM \cite{stipe98}. Although the principle is
similar, the conductance increases at the vibration energy in
IETS, while it decreases in PCS. Experimentally, a great advantage
is the short data acquisition time for a spectrum in our
experiment, 10\,s compared to 1--10\,h for IETS \cite{stipe98}.
This is attributed mainly to a lower shot noise level as result of
the lower resistance of the junction, and due to the quantum
suppression of shot noise for a single-channel contact
\cite{brom99}; the lower junction impedance also allows us to work
at a higher modulation frequency.

\begin{figure}[!t]
\includegraphics[width=100mm]{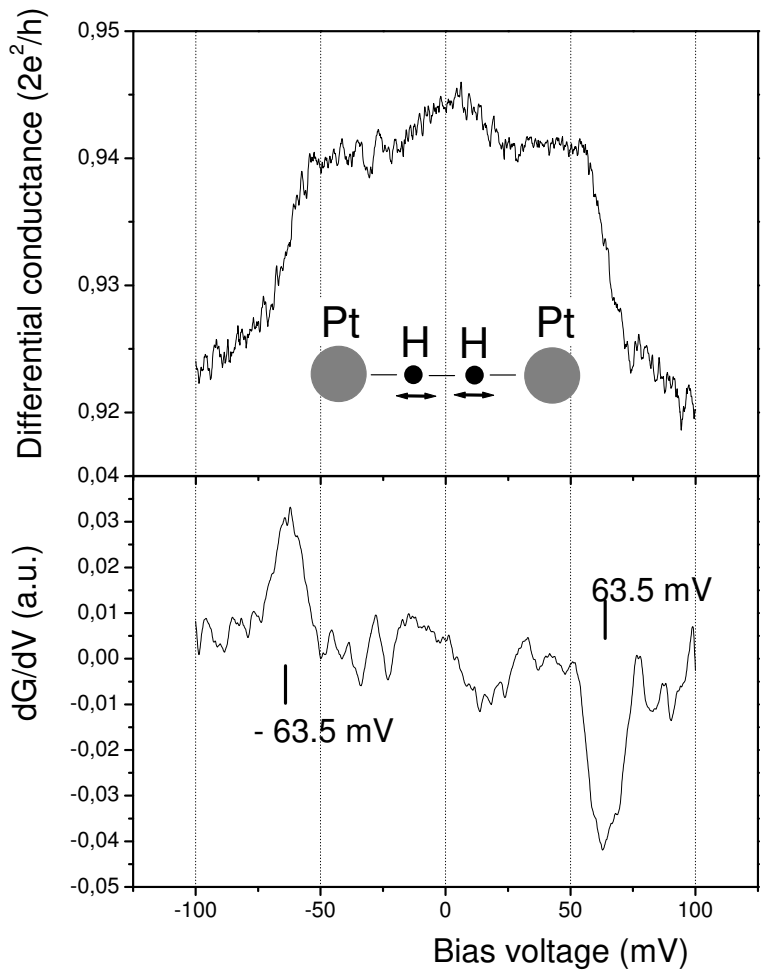}
\caption{Differential conductance (top) and its derivative
(bottom) for a Pt/H$_2$ contact taken at a conductance plateau
close to 1\,$G_0$. The differential conductance is recorded by a
lock-in amplifier using a modulation amplitude between 0.88 and
1.5\,mV$_{\rm rms}$ at 7\,kHz and a time constant of 10\,ms, and
the derivative is numerically calculated. A full spectrum is
recorded in 10\,s.\label{PCS-H2}}
\end{figure}

\begin{figure}[!t]
\includegraphics[width=100mm]{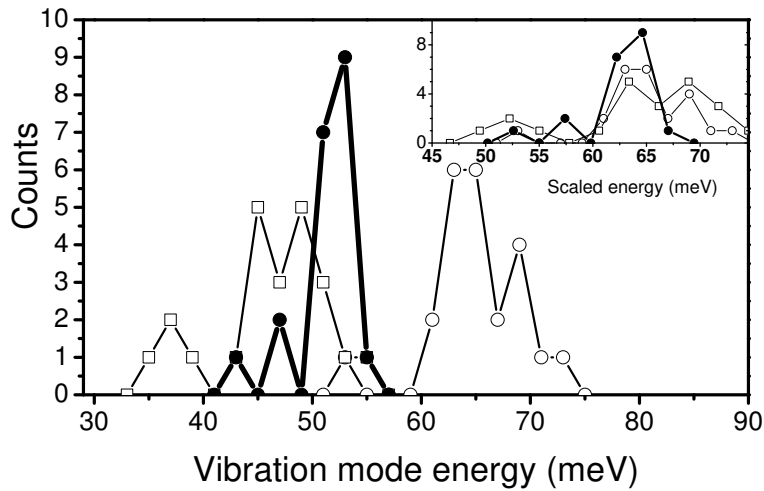}
\caption{Vibration mode energies obtained from point contact
spectra similar to that shown in Fig.\,\protect\ref{PCS-H2}, for
Pt/H$_2$ ($\circ$), Pt/D$_2$ ($\Box$), and  for Pt/HD ($\bullet$).
The vertical scale shows the number of spectra with energies
within a bin size of 2\,meV. The inset shows the same data with
the energy axis scaled by the factors expected for the isotope
shifts of the hydrogen molecule, $\omega_{H_{2}}/\omega_{D_{2}}
\sim \sqrt{m_{D_{2}}/m_{H_{2}}}=\sqrt{2}\simeq 1.414$ ($\Box$),
and $\omega_{H_{2}}/\omega_{HD} \sim \sqrt{m_{HD}/m_{H_{2}}} =
\sqrt{3/2} \simeq 1.225 $ ($\bullet$). \label{Isotopes}}
\end{figure}

Fig.\,\ref{PCS-H2} shows the differential conductance and its
derivative taken at a 1\,$G_0$-plateau for the Pt/H$_2$ system. We
find a pronounced single resonance at about 63.5\,mV,
symmetrically for both voltage polarities. The energy is much
higher than the typical phonon modes for metals, which are found
between 5 and 25\,mV \cite{khotkevich95}. The width of the
resonance is about 14\,mV and is much larger than expected from
the thermal and instrumental broadening. A similar large
'intrinsic width' has been observed in IETS \cite{stipe98} and it
probably results from the short lifetime of the molecular
vibration excitations due to the strong coupling to the metal. We
observe a modest variation in the position of the main signal
between different experiments, which is likely due to variations
in the bonding configuration of the hydrogen to the Pt electrodes.
The frequencies obtained from 23 spectra for Pt/H$_2$ are shown by
the open circles in Fig.\,\ref{Isotopes}, having a mean value of
64\,mV and a standard deviation of 4\,mV.

In order to test the interpretation of the observed resonance we
repeated the experiment using the isotopes D$_2$, and HD. From 23
spectra for Pt/D$_2$ we obtain a distribution of energies shown by
open squares in Fig.\,\ref{Isotopes}, being centred at 47\,mV,
while 20 spectra obtained for Pt/HD (bullets) are found to be
centred at 51\,mV . The inset to Fig.\,\ref{Isotopes} shows the
same distributions with energies scaled by the expected ratios for
the vibration energies $\omega_{H_{2}}/\omega_{D_{2}} \sim
\sqrt{m_{D_{2}}/m_{H_{2}}}=\sqrt{2}\simeq 1.414$, and
$\omega_{H_{2}}/\omega_{HD} \sim \sqrt{m_{HD}/m_{H_{2}}} =
\sqrt{3/2} \simeq 1.225 $, confirming our interpretation of the
conduction through a hydrogen molecule. Note in particular that
this excludes the possibility of conductance through a hydrogen
{\it atom}, since this would have resulted in a two-peak
distribution of frequencies for HD.

\begin{figure}[!t]
\includegraphics[width=140mm]{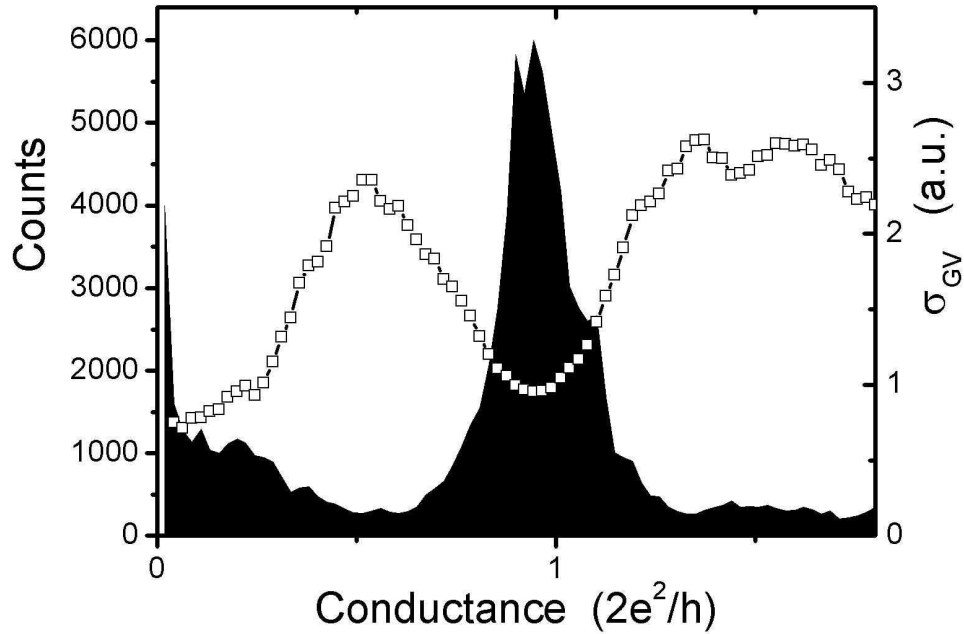}
\caption{Conductance histogram (black, left axis) and rms
amplitude of the conductance fluctuations   ($\Box$, right axis)
for a Pt/H$_2$ sample obtained using 2000 cycles of contact
breaking. The conductance and its derivative were measured with
two parallel lock-in amplifiers, detecting the frequencies $f$ and
$2f$ , with 140\,mV bias voltage and 20\,mV modulation amplitude.
The derivative signal is used to calculate the average of the
conductance fluctuations, $\sigma_{\rm GV}$, and each of the
points is obtained from the data belonging to one bin of the
histogram. \label{sigmaGV}}
\end{figure}

Experimental information on the number of conductance modes can be
obtained from the fluctuations in $G_{\rm d}(V)$  as a function of
voltage by measuring their root-mean square amplitude $\sigma_{\rm
GV}$ following the method described in
Ref.\,\onlinecite{ludoph00a}. For a contact with a single
conductance channel with transmission probability $T<1$ the
dominant contribution to the fluctuations results from
interference of partial waves reflected at the contact itself and
those reflected on defects nearby. The wavelength of the electrons
changes as a function of the bias voltage, producing random
variations in the interference and thus in the conductance. For
$T=1$ the reflection at the contact vanishes resulting in a
suppression of the fluctuations. Fig.\,\ref{sigmaGV} shows that
the peak in the conductance histogram at 0.95\,$G_0$ coincides
with a pronounced minimum in $\sigma_{\rm GV}$. From the finite
value of at the minimum \cite{ludoph00a} we extract a value for
the transmission $T = 0.97 \pm 0.01$, confirming that the
conductance through the molecule is almost entirely carried by one
channel. This finding also excludes other configurations for which
the conductance would be carried through several parallel channels
and confirms that we have only a single molecule.

Hydrogen is known to bind strongly to a Pt surface, and Pt
surfaces catalyse hydrogen dissociation. Little is known about the
catalytic activity of this system at 4.2\,K, but it is likely that
a small energy barrier is present that prevents H$_2$ dissociation
under these conditions. In the experiment we cannot rule out the
formation of contacts with atomic hydrogen, but the ones that we
have been able to fully analyse have a molecular bridge. We have
verified the stability of the H$_2$-bridge configuration using
density functional calculations employing the Gaussian 98 program
\cite{frisch98} with the SDD relativistic effective core basis set
\cite{andrae90} and the B3LYP functional \cite{becke93}.
Calculations were performed starting with a linear chain of four
Pt atoms, and the results were verified to be insensitive to
adding more atoms to the length of the Pt chain. Performing
first-principles molecular dynamics we find that a H$_2$ molecule
bonded to the side of a Pt chain spontaneously moves as a whole
into the chain when the bonds between the Pt atoms are being
stretched. The energy gain is 2.0\,eV, as compared to free H$_2$
and two Pt$_2$ chain fragments at infinity. The binding energy of
the Pt atoms in the chain is 2.75\,eV per bond. H$_2$ will
therefore never move spontaneously into the chain but first
requires an external force to stretch a Pt-Pt bond, which is a way
to supply chemical energy to a single bond. In order to obtain the
longitudinal vibrational modes of H$_2$ in a Pt chain, we have
explicitly calculated the potential energy curves. For the center
of mass motion of H$_2$ we obtain an excitation energy of
61.5\,meV, using equilibrium bond distances in a linear
arrangement of 0.08\,nm and 0.21\,nm for H-H and Pt-H,
respectively. This vibration mode fits very closely the observed
excitation energy. The internal vibration mode of H$_2$ is too
high in energy ($\sim430$\,meV) to be observed in the experiment
since the contact becomes unstable at bias voltages above about
200\,mV.

The conductance of the hydrogen bridge was calculated using a
model of a chain of Pt-H-H-Pt, sandwiched between two jellium bulk
electrodes. We refer to Ref.\,\onlinecite{lang95} for more details
on the computation method. For the equilibrium bond distances we
find a conductance of 0.9\,$G_0$, in close agreement with
experiment. We have not attempted to obtain the eigenchannel
decomposition of the conductance.

The conduction through the molecule involves mainly the H$_2$
antibonding states, but the hybridisation with the Pt metal states
is strong enough to largely fill the HOMO-LUMO gap. It is
surprising that the closed-shell configuration of H$_2$ permits
such strong bonding with Pt, while the molecular character is
largely conserved. The latter is evidenced by the calculated H-H
bond distance, which is close to that of the free molecule, and
the fact that in the simulations upon further stretching the
bridge finally breaks at the Pt-H bond. A Pt one-atom contact is
expected to have five conductance channels due to the partially
occupied $d$-orbitals. It appears that the insertion of a hydrogen
molecule has the effect of filtering out a single one from these
five channels, with nearly perfect transmission.

It will be interesting to attempt to extend our technique to more
complex molecules with built-in functional groups. Although most
organic molecules are expected to have a conductance many orders
of magnitude below the quantum unit, our experiments confirm
\cite{lang01} that full transmission of a single channel is
possible when the coupling to the leads is sufficiently strong.
Very recently two groups have demonstrated conductance through
single metalorganic molecules \cite{park02,liang02}, for which the
charge-state of the metal ions could even be controlled by a gate
electrode. In stead of using mechanical adjustment of the contact
size as in our experiments the size of the metal contacts to the
molecule was reduced exploiting electromigration. This further
illustrates that we are rapidly developing the tools to study and
control electron transport at single-molecule level.

We gratefully acknowledge discussions with Alfredo Levy Yeyati and
S{\o}ren Kynde Nielsen and we thank David Bakker and Marcel
Pohlkamp for valuable assistance in the experiments. CU and YN
have been supported by European Community Marie Curie Fellowships.

\bibliography{F:/User/Papers/my_QPC}

\begin{thebibliography}{32}
\expandafter\ifx\csname natexlab\endcsname\relax\def\natexlab#1{#1}\fi
\expandafter\ifx\csname bibnamefont\endcsname\relax
  \def\bibnamefont#1{#1}\fi
\expandafter\ifx\csname bibfnamefont\endcsname\relax
  \def\bibfnamefont#1{#1}\fi
\expandafter\ifx\csname citenamefont\endcsname\relax
  \def\citenamefont#1{#1}\fi
\expandafter\ifx\csname url\endcsname\relax
  \def\url#1{\texttt{#1}}\fi
\expandafter\ifx\csname urlprefix\endcsname\relax\def\urlprefix{URL }\fi
\providecommand{\bibinfo}[2]{#2}
\providecommand{\eprint}[2][]{\url{#2}}

\bibitem[{\citenamefont{Aviram and Ratner}(1998)}]{aviram98}
\bibinfo{editor}{\bibfnamefont{A.}~\bibnamefont{Aviram}} \bibnamefont{and}
  \bibinfo{editor}{\bibfnamefont{M.}~\bibnamefont{Ratner}}, eds.,
  \emph{\bibinfo{title}{Molecular electronics: science and technology}}
  (\bibinfo{publisher}{Annals of the New York Academy of Sciences},
  \bibinfo{address}{New York}, \bibinfo{year}{1998}).

\bibitem[{\citenamefont{Langlais et~al.}(1999)\citenamefont{Langlais,
  Schlittler, Tang, Gourdon, Joachim, and Gimzewski}}]{langlais99}
\bibinfo{author}{\bibfnamefont{V.}~\bibnamefont{Langlais}},
  \bibinfo{author}{\bibfnamefont{R.}~\bibnamefont{Schlittler}},
  \bibinfo{author}{\bibfnamefont{H.}~\bibnamefont{Tang}},
  \bibinfo{author}{\bibfnamefont{A.}~\bibnamefont{Gourdon}},
  \bibinfo{author}{\bibfnamefont{C.}~\bibnamefont{Joachim}}, \bibnamefont{and}
  \bibinfo{author}{\bibfnamefont{J.}~\bibnamefont{Gimzewski}},
  \bibinfo{journal}{Phys. Rev. Lett.} \textbf{\bibinfo{volume}{83}},
  \bibinfo{pages}{2809} (\bibinfo{year}{1999}).

\bibitem[{\citenamefont{Gao et~al.}(2000)\citenamefont{Gao, Sohlberg, Xue,
  Chen, Hou, Ma, Fang, Pang, and Pennycook}}]{gao00}
\bibinfo{author}{\bibfnamefont{H.}~\bibnamefont{Gao}},
  \bibinfo{author}{\bibfnamefont{K.}~\bibnamefont{Sohlberg}},
  \bibinfo{author}{\bibfnamefont{Z.}~\bibnamefont{Xue}},
  \bibinfo{author}{\bibfnamefont{H.}~\bibnamefont{Chen}},
  \bibinfo{author}{\bibfnamefont{S.}~\bibnamefont{Hou}},
  \bibinfo{author}{\bibfnamefont{L.}~\bibnamefont{Ma}},
  \bibinfo{author}{\bibfnamefont{X.}~\bibnamefont{Fang}},
  \bibinfo{author}{\bibfnamefont{S.}~\bibnamefont{Pang}}, \bibnamefont{and}
  \bibinfo{author}{\bibfnamefont{S.}~\bibnamefont{Pennycook}},
  \bibinfo{journal}{Phys. Rev. Lett.} \textbf{\bibinfo{volume}{84}},
  \bibinfo{pages}{1780} (\bibinfo{year}{2000}).

\bibitem[{\citenamefont{Collier et~al.}(1999)\citenamefont{Collier, Wong,
  Belohradsk{\'y}, Raymo, Stoddart, Kuekes, Williams, and Heath}}]{collier99}
\bibinfo{author}{\bibfnamefont{C.}~\bibnamefont{Collier}},
  \bibinfo{author}{\bibfnamefont{E.}~\bibnamefont{Wong}},
  \bibinfo{author}{\bibfnamefont{M.}~\bibnamefont{Belohradsk{\'y}}},
  \bibinfo{author}{\bibfnamefont{F.}~\bibnamefont{Raymo}},
  \bibinfo{author}{\bibfnamefont{J.}~\bibnamefont{Stoddart}},
  \bibinfo{author}{\bibfnamefont{P.}~\bibnamefont{Kuekes}},
  \bibinfo{author}{\bibfnamefont{R.}~\bibnamefont{Williams}}, \bibnamefont{and}
  \bibinfo{author}{\bibfnamefont{J.}~\bibnamefont{Heath}},
  \bibinfo{journal}{Science} \textbf{\bibinfo{volume}{285}},
  \bibinfo{pages}{391} (\bibinfo{year}{1999}).

\bibitem[{\citenamefont{Reed et~al.}(2001)\citenamefont{Reed, Chen, Rawlett,
  Price, and Tour}}]{reed01}
\bibinfo{author}{\bibfnamefont{M.}~\bibnamefont{Reed}},
  \bibinfo{author}{\bibfnamefont{J.}~\bibnamefont{Chen}},
  \bibinfo{author}{\bibfnamefont{A.}~\bibnamefont{Rawlett}},
  \bibinfo{author}{\bibfnamefont{D.}~\bibnamefont{Price}}, \bibnamefont{and}
  \bibinfo{author}{\bibfnamefont{J.}~\bibnamefont{Tour}},
  \bibinfo{journal}{Appl. Phys. Lett.} \textbf{\bibinfo{volume}{78}},
  \bibinfo{pages}{3735} (\bibinfo{year}{2001}).

\bibitem[{\citenamefont{Metzger and Cava}(1998)}]{metzger98}
\bibinfo{author}{\bibfnamefont{R.}~\bibnamefont{Metzger}} \bibnamefont{and}
  \bibinfo{author}{\bibfnamefont{M.}~\bibnamefont{Cava}},
  \bibinfo{journal}{Ann. New York Acad. Sci.} \textbf{\bibinfo{volume}{852}},
  \bibinfo{pages}{95} (\bibinfo{year}{1998}).

\bibitem[{\citenamefont{Chen et~al.}(1999)\citenamefont{Chen, Reed, Rawlett,
  and Tour}}]{chen-j99}
\bibinfo{author}{\bibfnamefont{J.}~\bibnamefont{Chen}},
  \bibinfo{author}{\bibfnamefont{M.}~\bibnamefont{Reed}},
  \bibinfo{author}{\bibfnamefont{A.}~\bibnamefont{Rawlett}}, \bibnamefont{and}
  \bibinfo{author}{\bibfnamefont{J.}~\bibnamefont{Tour}},
  \bibinfo{journal}{Science} \textbf{\bibinfo{volume}{286}},
  \bibinfo{pages}{1550} (\bibinfo{year}{1999}).

\bibitem[{\citenamefont{Reed et~al.}(1997)\citenamefont{Reed, Zhou, Muller,
  Burgin, and Tour}}]{reed97}
\bibinfo{author}{\bibfnamefont{M.}~\bibnamefont{Reed}},
  \bibinfo{author}{\bibfnamefont{C.}~\bibnamefont{Zhou}},
  \bibinfo{author}{\bibfnamefont{C.}~\bibnamefont{Muller}},
  \bibinfo{author}{\bibfnamefont{T.}~\bibnamefont{Burgin}}, \bibnamefont{and}
  \bibinfo{author}{\bibfnamefont{J.}~\bibnamefont{Tour}},
  \bibinfo{journal}{Science} \textbf{\bibinfo{volume}{278}},
  \bibinfo{pages}{252} (\bibinfo{year}{1997}).

\bibitem[{\citenamefont{Kergueris et~al.}(1999)\citenamefont{Kergueris,
  Bourgoin, Palacin, Esteve, Urbina, Magoga, and Joachim}}]{kergueris99}
\bibinfo{author}{\bibfnamefont{C.}~\bibnamefont{Kergueris}},
  \bibinfo{author}{\bibfnamefont{J.-P.} \bibnamefont{Bourgoin}},
  \bibinfo{author}{\bibfnamefont{S.}~\bibnamefont{Palacin}},
  \bibinfo{author}{\bibfnamefont{D.}~\bibnamefont{Esteve}},
  \bibinfo{author}{\bibfnamefont{C.}~\bibnamefont{Urbina}},
  \bibinfo{author}{\bibfnamefont{M.}~\bibnamefont{Magoga}}, \bibnamefont{and}
  \bibinfo{author}{\bibfnamefont{C.}~\bibnamefont{Joachim}},
  \bibinfo{journal}{Phys. Rev. B} \textbf{\bibinfo{volume}{59}},
  \bibinfo{pages}{12505} (\bibinfo{year}{1999}).

\bibitem[{\citenamefont{Reichert et~al.}(2002)\citenamefont{Reichert, Ochs,
  Beckmann, Weber, Mayor, and von L{\"o}hneisen}}]{reichert02}
\bibinfo{author}{\bibfnamefont{J.}~\bibnamefont{Reichert}},
  \bibinfo{author}{\bibfnamefont{R.}~\bibnamefont{Ochs}},
  \bibinfo{author}{\bibfnamefont{D.}~\bibnamefont{Beckmann}},
  \bibinfo{author}{\bibfnamefont{H.}~\bibnamefont{Weber}},
  \bibinfo{author}{\bibfnamefont{M.}~\bibnamefont{Mayor}}, \bibnamefont{and}
  \bibinfo{author}{\bibfnamefont{H.}~\bibnamefont{von L{\"o}hneisen}},
  \bibinfo{journal}{Phys. Rev. Lett.} \textbf{\bibinfo{volume}{88}},
  \bibinfo{pages}{176804} (\bibinfo{year}{2002}).

\bibitem[{\citenamefont{Emberly and Kirczenow}(2001)}]{emberly01}
\bibinfo{author}{\bibfnamefont{E.}~\bibnamefont{Emberly}} \bibnamefont{and}
  \bibinfo{author}{\bibfnamefont{G.}~\bibnamefont{Kirczenow}},
  \bibinfo{journal}{Phys. Rev. Lett.} \textbf{\bibinfo{volume}{87}},
  \bibinfo{pages}{269701} (\bibinfo{year}{2001}).

\bibitem[{\citenamefont{Muller et~al.}(1992)\citenamefont{Muller, van
  Ruitenbeek, and de~Jongh}}]{muller92a}
\bibinfo{author}{\bibfnamefont{C.}~\bibnamefont{Muller}},
  \bibinfo{author}{\bibfnamefont{J.}~\bibnamefont{van Ruitenbeek}},
  \bibnamefont{and} \bibinfo{author}{\bibfnamefont{L.}~\bibnamefont{de~Jongh}},
  \bibinfo{journal}{Physica C} \textbf{\bibinfo{volume}{191}},
  \bibinfo{pages}{485} (\bibinfo{year}{1992}).

\bibitem[{\citenamefont{van Ruitenbeek}(1997)}]{ruitenbeek97a}
\bibinfo{author}{\bibfnamefont{J.}~\bibnamefont{van Ruitenbeek}}, in
  \emph{\bibinfo{booktitle}{Mesoscopic Electron Transport}}, edited by
  \bibinfo{editor}{\bibfnamefont{L.}~\bibnamefont{Sohn}},
  \bibinfo{editor}{\bibfnamefont{L.}~\bibnamefont{Kouwenhoven}},
  \bibnamefont{and} \bibinfo{editor}{\bibfnamefont{G.}~\bibnamefont{Sch{\"o}n}}
  (\bibinfo{publisher}{Kluwer Academic Publishers},
  \bibinfo{address}{Dordrecht}, \bibinfo{year}{1997}), vol.
  \bibinfo{volume}{345} of \emph{\bibinfo{series}{NATO-ASI Series E: Appl.
  Sci.}}, pp. \bibinfo{pages}{549--579}.

\bibitem[{\citenamefont{Rubio et~al.}(1996)\citenamefont{Rubio, {Agra\"{\i}t},
  and Vieira}}]{rubio96}
\bibinfo{author}{\bibfnamefont{G.}~\bibnamefont{Rubio}},
  \bibinfo{author}{\bibfnamefont{N.}~\bibnamefont{{Agra\"{\i}t}}},
  \bibnamefont{and} \bibinfo{author}{\bibfnamefont{S.}~\bibnamefont{Vieira}},
  \bibinfo{journal}{Phys. Rev. Lett.} \textbf{\bibinfo{volume}{76}},
  \bibinfo{pages}{2302} (\bibinfo{year}{1996}).

\bibitem[{\citenamefont{Agra{\"\i}t
  et~al.}(2002{\natexlab{a}})\citenamefont{Agra{\"\i}t, Yeyati, and van
  Ruitenbeek}}]{agrait02b}
\bibinfo{author}{\bibfnamefont{N.}~\bibnamefont{Agra{\"\i}t}},
  \bibinfo{author}{\bibfnamefont{A.~L.} \bibnamefont{Yeyati}},
  \bibnamefont{and} \bibinfo{author}{\bibfnamefont{J.}~\bibnamefont{van
  Ruitenbeek}}, \bibinfo{journal}{Phys. Rep.}
  (\bibinfo{year}{2002}{\natexlab{a}}), \bibinfo{note}{submitted; preprint
  http://xxx.lanl.gov/abs/cond-mat/0208239}.

\bibitem[{\citenamefont{Yanson}(1974)}]{yanson74}
\bibinfo{author}{\bibfnamefont{I.}~\bibnamefont{Yanson}}, \bibinfo{journal}{Zh.
  Eksp. Teor. Fiz.} \textbf{\bibinfo{volume}{66}}, \bibinfo{pages}{1035}
  (\bibinfo{year}{1974}), \bibinfo{note}{[Sov. Phys.-JETP {\bf 39} (1974)
  506--513]}.

\bibitem[{\citenamefont{Jansen et~al.}(1980)\citenamefont{Jansen, van Gelder,
  and Wyder}}]{jansen80}
\bibinfo{author}{\bibfnamefont{A.}~\bibnamefont{Jansen}},
  \bibinfo{author}{\bibfnamefont{A.}~\bibnamefont{van Gelder}},
  \bibnamefont{and} \bibinfo{author}{\bibfnamefont{P.}~\bibnamefont{Wyder}},
  \bibinfo{journal}{J. Phys. C: Solid St. Phys.} \textbf{\bibinfo{volume}{13}},
  \bibinfo{pages}{6073} (\bibinfo{year}{1980}).

\bibitem[{\citenamefont{Untiedt et~al.}(2000)\citenamefont{Untiedt, {Rubio
  Bollinger}, Vieira, and Agra{\"\i}t}}]{untiedt00}
\bibinfo{author}{\bibfnamefont{C.}~\bibnamefont{Untiedt}},
  \bibinfo{author}{\bibfnamefont{G.}~\bibnamefont{{Rubio Bollinger}}},
  \bibinfo{author}{\bibfnamefont{S.}~\bibnamefont{Vieira}}, \bibnamefont{and}
  \bibinfo{author}{\bibfnamefont{N.}~\bibnamefont{Agra{\"\i}t}},
  \bibinfo{journal}{Phys. Rev. B} \textbf{\bibinfo{volume}{62}},
  \bibinfo{pages}{9962} (\bibinfo{year}{2000}).

\bibitem[{\citenamefont{Agra{\"\i}t
  et~al.}(2002{\natexlab{b}})\citenamefont{Agra{\"\i}t, Untiedt,
  Rubio-Bollinger, and Vieira}}]{agrait02a}
\bibinfo{author}{\bibfnamefont{N.}~\bibnamefont{Agra{\"\i}t}},
  \bibinfo{author}{\bibfnamefont{C.}~\bibnamefont{Untiedt}},
  \bibinfo{author}{\bibfnamefont{G.}~\bibnamefont{Rubio-Bollinger}},
  \bibnamefont{and} \bibinfo{author}{\bibfnamefont{S.}~\bibnamefont{Vieira}},
  \bibinfo{journal}{Chem. Phys.} \textbf{\bibinfo{volume}{281}},
  \bibinfo{pages}{231} (\bibinfo{year}{2002}{\natexlab{b}}).

\bibitem[{\citenamefont{Bon{\v c}a and Trugman}(1995)}]{bonca95}
\bibinfo{author}{\bibfnamefont{J.}~\bibnamefont{Bon{\v c}a}} \bibnamefont{and}
  \bibinfo{author}{\bibfnamefont{S.}~\bibnamefont{Trugman}},
  \bibinfo{journal}{Phys. Rev. Lett.} \textbf{\bibinfo{volume}{75}},
  \bibinfo{pages}{2566} (\bibinfo{year}{1995}).

\bibitem[{\citenamefont{Emberly and Kircznow}(1999)}]{emberly99a}
\bibinfo{author}{\bibfnamefont{E.}~\bibnamefont{Emberly}} \bibnamefont{and}
  \bibinfo{author}{\bibfnamefont{G.}~\bibnamefont{Kircznow}},
  \bibinfo{journal}{Phys. Rev. B} \textbf{\bibinfo{volume}{61}},
  \bibinfo{pages}{5740} (\bibinfo{year}{1999}).

\bibitem[{\citenamefont{Stipe et~al.}(1998)\citenamefont{Stipe, Rezaei, and
  Ho}}]{stipe98}
\bibinfo{author}{\bibfnamefont{B.}~\bibnamefont{Stipe}},
  \bibinfo{author}{\bibfnamefont{M.}~\bibnamefont{Rezaei}}, \bibnamefont{and}
  \bibinfo{author}{\bibfnamefont{W.}~\bibnamefont{Ho}},
  \bibinfo{journal}{Science} \textbf{\bibinfo{volume}{280}},
  \bibinfo{pages}{1732} (\bibinfo{year}{1998}).

\bibitem[{\citenamefont{van~den Brom and van Ruitenbeek}(1999)}]{brom99}
\bibinfo{author}{\bibfnamefont{H.}~\bibnamefont{van~den Brom}}
  \bibnamefont{and} \bibinfo{author}{\bibfnamefont{J.}~\bibnamefont{van
  Ruitenbeek}}, \bibinfo{journal}{Phys. Rev. Lett.}
  \textbf{\bibinfo{volume}{82}}, \bibinfo{pages}{1526} (\bibinfo{year}{1999}).

\bibitem[{\citenamefont{Khotkevich and Yanson}(1995)}]{khotkevich95}
\bibinfo{author}{\bibfnamefont{A.}~\bibnamefont{Khotkevich}} \bibnamefont{and}
  \bibinfo{author}{\bibfnamefont{I.}~\bibnamefont{Yanson}},
  \emph{\bibinfo{title}{Atlas of point contact spectra of electron-phonon
  interactions in metals}} (\bibinfo{publisher}{Kluwer Academic Publishers},
  \bibinfo{address}{Dordrecht}, \bibinfo{year}{1995}).

\bibitem[{\citenamefont{Ludoph and van Ruitenbeek}(2000)}]{ludoph00a}
\bibinfo{author}{\bibfnamefont{B.}~\bibnamefont{Ludoph}} \bibnamefont{and}
  \bibinfo{author}{\bibfnamefont{J.}~\bibnamefont{van Ruitenbeek}},
  \bibinfo{journal}{Phys. Rev. B} \textbf{\bibinfo{volume}{61}},
  \bibinfo{pages}{2273} (\bibinfo{year}{2000}).

\bibitem[{\citenamefont{Frisch et~al.}()\citenamefont{Frisch, Trucks, Schlegel,
  Scuseria, Robb, Cheeseman, Zakrzewski, J.~A.~Montgomery, Stratmann, Burant
  et~al.}}]{frisch98}
\bibinfo{author}{\bibfnamefont{M.}~\bibnamefont{Frisch}},
  \bibinfo{author}{\bibfnamefont{G.~W.} \bibnamefont{Trucks}},
  \bibinfo{author}{\bibfnamefont{H.~B.} \bibnamefont{Schlegel}},
  \bibinfo{author}{\bibfnamefont{G.~E.} \bibnamefont{Scuseria}},
  \bibinfo{author}{\bibfnamefont{M.~A.} \bibnamefont{Robb}},
  \bibinfo{author}{\bibfnamefont{J.~R.} \bibnamefont{Cheeseman}},
  \bibinfo{author}{\bibfnamefont{V.~G.} \bibnamefont{Zakrzewski}},
  \bibinfo{author}{\bibfnamefont{J.}~\bibnamefont{J.~A.~Montgomery}},
  \bibinfo{author}{\bibfnamefont{R.~E.} \bibnamefont{Stratmann}},
  \bibinfo{author}{\bibfnamefont{J.~C.} \bibnamefont{Burant}},
  \bibnamefont{et~al.}, \bibinfo{note}{{G}aussian, Inc., Pittsburgh PA, 1998}.

\bibitem[{\citenamefont{Andrae et~al.}(1990)\citenamefont{Andrae,
  H{\"a}u{\ss}ermann, Dolg, Stoll, and Preu{\ss}}}]{andrae90}
\bibinfo{author}{\bibfnamefont{D.}~\bibnamefont{Andrae}},
  \bibinfo{author}{\bibfnamefont{U.}~\bibnamefont{H{\"a}u{\ss}ermann}},
  \bibinfo{author}{\bibfnamefont{M.}~\bibnamefont{Dolg}},
  \bibinfo{author}{\bibfnamefont{H.}~\bibnamefont{Stoll}}, \bibnamefont{and}
  \bibinfo{author}{\bibfnamefont{H.}~\bibnamefont{Preu{\ss}}},
  \bibinfo{journal}{Theor. Chim. Acta} \textbf{\bibinfo{volume}{77}},
  \bibinfo{pages}{123} (\bibinfo{year}{1990}).

\bibitem[{\citenamefont{Becke}(1993)}]{becke93}
\bibinfo{author}{\bibfnamefont{A.}~\bibnamefont{Becke}}, \bibinfo{journal}{J.
  Chem. Phys.} \textbf{\bibinfo{volume}{98}}, \bibinfo{pages}{5648}
  (\bibinfo{year}{1993}).

\bibitem[{\citenamefont{Lang}(1995)}]{lang95}
\bibinfo{author}{\bibfnamefont{N.}~\bibnamefont{Lang}}, \bibinfo{journal}{Phys.
  Rev. B} \textbf{\bibinfo{volume}{52}}, \bibinfo{pages}{5335}
  (\bibinfo{year}{1995}).

\bibitem[{\citenamefont{Lang and Avouris}(2001)}]{lang01}
\bibinfo{author}{\bibfnamefont{N.}~\bibnamefont{Lang}} \bibnamefont{and}
  \bibinfo{author}{\bibfnamefont{P.}~\bibnamefont{Avouris}},
  \bibinfo{journal}{Phys. Rev. B} \textbf{\bibinfo{volume}{64}},
  \bibinfo{pages}{125323} (\bibinfo{year}{2001}).

\bibitem[{\citenamefont{Park et~al.}(2002)\citenamefont{Park, Pasupathy,
  Goldsmith, Chang, Yaish, Petta, Rinkoski, Sethna, Abru{\~n}a, Mc{E}uen
  et~al.}}]{park02}
\bibinfo{author}{\bibfnamefont{J.}~\bibnamefont{Park}},
  \bibinfo{author}{\bibfnamefont{A.}~\bibnamefont{Pasupathy}},
  \bibinfo{author}{\bibfnamefont{J.}~\bibnamefont{Goldsmith}},
  \bibinfo{author}{\bibfnamefont{C.}~\bibnamefont{Chang}},
  \bibinfo{author}{\bibfnamefont{Y.}~\bibnamefont{Yaish}},
  \bibinfo{author}{\bibfnamefont{J.}~\bibnamefont{Petta}},
  \bibinfo{author}{\bibfnamefont{M.}~\bibnamefont{Rinkoski}},
  \bibinfo{author}{\bibfnamefont{J.}~\bibnamefont{Sethna}},
  \bibinfo{author}{\bibfnamefont{H.}~\bibnamefont{Abru{\~n}a}},
  \bibinfo{author}{\bibfnamefont{P.}~\bibnamefont{Mc{E}uen}},
  \bibnamefont{et~al.}, \bibinfo{journal}{Nature}
  \textbf{\bibinfo{volume}{417}}, \bibinfo{pages}{722} (\bibinfo{year}{2002}).

\bibitem[{\citenamefont{Liang et~al.}(2002)\citenamefont{Liang, Shores,
  Bockrath, Long, and Park}}]{liang02}
\bibinfo{author}{\bibfnamefont{W.}~\bibnamefont{Liang}},
  \bibinfo{author}{\bibfnamefont{M.}~\bibnamefont{Shores}},
  \bibinfo{author}{\bibfnamefont{M.}~\bibnamefont{Bockrath}},
  \bibinfo{author}{\bibfnamefont{J.}~\bibnamefont{Long}}, \bibnamefont{and}
  \bibinfo{author}{\bibfnamefont{H.}~\bibnamefont{Park}},
  \bibinfo{journal}{Nature} \textbf{\bibinfo{volume}{417}},
  \bibinfo{pages}{725} (\bibinfo{year}{2002}).

\end{thebibliography}

\end{document}